\documentclass[aps,twocolumn,amsmath,amssymb,eqsecnum,prb]{revtex4}
\usepackage{graphicx}      % Include figure files
\usepackage{dcolumn}       % Align table columns on decimal point

\usepackage{color}
\usepackage{longtable}

\newcommand{\beq}{\begin{equation}}
\newcommand{\eeq}{\end{equation}}
\newcommand{\beqn}{\begin{eqnarray}}
\newcommand{\eeqn}{\end{eqnarray}}

\newcommand\beqa{\begin{eqnarray}}
\newcommand\eeqa{\end{eqnarray}}

\newcommand{\dd}{\text{d}}

\newcommand{\an}{\lambda}

\begin{document}

\title{Virial series for fluids of hard hyperspheres in odd dimensions}
\author{Ren\'e D. Rohrmann}
\email{rohr@oac.uncor.edu}

\affiliation{Observatorio Astron\'omico, Universidad Nacional de
C\'ordoba, Laprida 854, X5000BGR C\'ordoba, Argentina}
\author{Miguel Robles}
\email{mrp@cie.unam.mx}
\homepage{http://xml.cie.unam.mx/xml/tc/ft/mrp/}

\author{Mariano L\'opez de Haro}
\email{malopez@servidor.unam.mx}
\homepage{http://xml.cie.unam.mx/xml/tc/ft/mlh/}
\affiliation{Centro
de Investigaci\'on en Energ\'{\i}a, Universidad Nacional Aut\'onoma
de M\'exico (U.N.A.M.), Temixco, Morelos 62580, M{e}xico}
\author{Andr\'es Santos}
\email{andres@unex.es}
\homepage{http://www.unex.es/fisteor/andres/}
\affiliation{Departamento de F\'{\i}sica, Universidad de
Extremadura, E-06071 Badajoz, Spain}

%\maketitle
\date{\today}
\begin{abstract}
A recently derived method [R. D. Rohrmann and A. Santos, Phys. Rev.
E. {\bf 76}, 051202 (2007)] to obtain the exact solution of the
Percus--Yevick equation for a fluid of hard spheres in (odd) $d$
dimensions is used to investigate the convergence properties of the
resulting virial series. This is done both for the virial and
compressibility routes, in which the virial coefficients $B_j$ are
expressed in terms of the solution of a set of $(d-1)/2$ coupled
algebraic equations which become nonlinear for $d \geq 5$. Results
have been derived up to $d=13$. A confirmation of the alternating
character of the series for $d\geq 5$, due to the existence of a
branch point on the negative real axis, is found and the radius of
convergence is explicitly determined for each dimension. The
resulting scaled density per dimension $2 \eta^{1/d}$, where $\eta$
is the packing fraction, is wholly consistent with the limiting
value of $1$ for $d \to \infty$. Finally, the values for $B_j$
predicted by the virial and compressibility routes in the
Percus--Yevick approximation are compared with the known exact
values [N. Clisby and B. M. McCoy, J. Stat. Phys. {\bf 122}, 15
(2006)].
\end{abstract}

%\pacs{61.20.Gy, 61.20.Ne, 05.20.Jj, 51.30.+i} % PACS, the Physics and Astronomy
                                    % Classification Scheme.
      %02.50.-r Probability theory, stochastic processes and statistics
      %05.20.Jj Statistical mechanics of classical fluids
      %34.20.Cf Interatomic potentials and forces
      %51.30.+i Thermodynamic properties, equations of state
      %52.25.Jm Ionization of plasmas
      %61.20.Gy    Theory and models of liquid structure
      %61.20.Ne    Structure of simple liquids

\maketitle
\section{Introduction}
\label{sec1}

Interest in studying fluids of hard spheres in $d$ dimensions goes
back at least four
decades\cite{RH64,FI81,LB82,J82,L84,MT84,BMS85,FRW85,L86,FRW86,KF86,CB86,FP87,WRF87,BC87,R87,BR88,EF88,SMS89,ASV89,LM90,SM90,
GGM90,MSAV91,GGM91,LZKH91,CFP91,GGM92} and has recently experienced
a new
boom.\cite{SYH99,MP99,FP99,VMN99,BMC99,YFP00,S00,PS00,YSH00,SYH01,GAH01,EAGB02,SYH02,FSL02,CM03,BMV04,Ly05,BMV05,CM05,CM06,
RLdeHS04,SH05,L05,BWK05,BW05,LB06,SDST06,BW07,RLdeHS07,RS07,BCW08,A-BKV08}
The evaluation of virial coefficients, the derivation of the
equation of state for these systems, or the determination of the
radius of convergence of the virial series are among the issues that
have been examined, but many questions related to these issues and
others are still open. Providing answers to these questions may shed
light on similar issues related to real fluids and therefore efforts
in this direction are called for.

The Percus--Yevick (PY) {theory}\cite{PY58} is one of the classic
integral-equation approximations of liquid-state
theory.\cite{HMcD06} Apart from yielding the correct expression for
the radial distribution function {(rdf)} to first order in the
density (and hence also the correct second and third virial
coefficients), its key role in the theory of simple liquids was
recognized very early because the resulting integral equation is
analytically solvable for the important case of the hard-sphere
fluid ($d=3$).\cite{W63,T63} Further, the approximation {provides
the exact rdf (although not the exact cavity function inside the
core\cite{MS06})} for hard rods ($d=1$)  and has been proven to
yield exactly solvable equations in odd $d$
dimensions.\cite{FI81,L84} In fact, explicit analytical solutions
for $d=5$ and $d=7$ have been
obtained\cite{FI81,L84,RLdeHS04,RLdeHS07} and, rather recently,
numerical solutions for $d=9$ and $d=11$ have been
reported.\cite{RS07} These latter have been derived using an
alternative method to the one originally employed by
Leutheusser\cite{L84} which, amongst other things, allows one to
 obtain the virial coefficients and the equation of
state both from the virial and compressibility routes in a rather
straightforward procedure. Also very recently, Adda-Bedia \emph{et
al.}\cite{A-BKV08} have been able to reduce the PY equation for hard
disks ($d=2$) to a set of simple integral equations which they then
solve numerically. An interesting aspect of this work is their claim
that the method may be generalized to any even dimension.

Due to the limited availability of virial coefficients and to the
fact that the virial series for high densities relevant to the
fluid phase seems to be in general either divergent or slowly
convergent, various series convergence accelerating methods, such
as Pad\'e or Levin approximants, have been used to derive
approximate equations of state for $d$-dimensional hard-sphere
fluids. However, the radius of convergence of the virial series is
not known in general, and hence the usefulness of such approximate
equations of state is limited by the uncertainty of its range of
applicability. For $d\leq 3$ all known virial coefficients turn
out to be positive but since negative virial coefficients appear
for $d \geq 4$, the question arises as to whether some higher
order virial coefficients both for $d=2$ and $d=3$ might
eventually become negative.

The main purpose of this paper is to examine the question of
convergence of the virial series for fluids of $d$-dimensional hard
spheres by looking at approximate theories where both the virial
coefficients and the equation of state are known. We will use the
procedure introduced in Ref.\ \onlinecite{RS07} to derive the
equation of state of the system and the values for the virial
coefficients taking both the virial and compressibility routes, all
within the PY approximation. In particular, we will investigate the
convergence properties of the virial series of the equations of
state stemming out of both routes for $d=5,7,9,11$, and $13$ and
will show that the radius of convergence is related to a branch
point on the negative real axis. Moreover, we will compare the PY
virial coefficients with the exact values available in the
literature. This comparison suggests that, as the dimensionality
increases, the true radius of convergence tends to the value
predicted by the PY theory. We will also examine the performance of
the compressibility factors obtained from both routes with the
corresponding simulation data and will show that the virial series
truncated just before the first negative term provides an excellent
approximation to the equation of state of the fluid phase.

The paper is organized as follows. In the next section and in
order to make the development self-contained, we provide a brief
description of the so-called Rational Function Approximation (RFA)
approach leading to the PY approximation for fluids of hard
hyperspheres in odd dimensions. Section \ref{sect3} presents the
numerical results of our calculations (together with a comparison
with simulation data) for the PY compressibility factors derived
via the virial and compressibility routes, respectively, as well
as the analysis of the behavior of the PY virial coefficients
obtained from the same routes to the equation of state both with
respect to the convergence properties of the virial series and
with respect to the exact known values. The paper is closed in
Sect.\ \ref{sect4} with some discussion and concluding remarks.

\section{The PY theory}
\label{sect2}

In this section we provide an outline of the main steps leading to
the PY approximation for the thermodynamic and structural properties
of hard hypersphere fluids in odd dimensions. Instead of following
the original derivation by Leutheusser,\cite{L84} we will use the
RFA method introduced in Ref.\ \onlinecite{RS07} which the reader is
urged to consult for further details.

\subsection{General background}
We begin by recalling that the static structure factor of a
$d$-dimensional hard-sphere fluid can be written as
\beq \label{SG} S(k)=1+\rho \frac{(2\pi)^{(d-1)/2}}{k^{d-2}} i
\left[ G(ik)-G(-ik) \right], \eeq
where $\rho$ is the particle number density and $G(s)$ is a
Laplace-space functional given by\cite{RS07}
\beq
\label{GH}
G(s) =\int_0^\infty \dd r\,r g(r) \theta_n(sr)e^{-sr}.
\eeq
Here $g(r)$ is the rdf and $\theta_n(t)$ is the reverse Bessel
polynomial of degree $n\equiv(d-3)/2$. From these structural
functions one may obtain thermodynamic quantities such as the
compressibility factor $Z\equiv p/\rho k_B T $ and the isothermal
susceptibility $\chi \equiv k_B T (\partial \rho/\partial p)_T$,
both defined in terms of the temperature $T$, the pressure $p$ and
the number density $\rho$ ($k_B$ being the Boltzmann constant). The
actual relationships read
\beq
\label{Zg}
Z = 1 + 2^{d-1} \eta  g(\sigma^+)
\eeq
and
\beq \label{chiS} \chi = S(0), \eeq
where $\sigma$ is the diameter of the particles and $\eta$ is the
packing fraction given by
\beq \label{eta} \eta=v_d \,\rho \,\sigma^d, \quad
v_d=\frac{(\pi/4)^{d/2}}{\Gamma(1+d/2)}, \eeq
$v_d$ being the volume of a $d$-sphere of unit diameter.
Henceforth without loss of generality we will set $\sigma=1$.

In hard-particle systems the temperature does not play any
relevant role on the structural properties here introduced.
Moreover, the thermodynamic state of such fluids can be
characterized by a variable alone, say the density or the packing
fraction. Therefore, taking into account the thermodynamic
relation
\beq \label{chiZ} \chi^{-1}=\frac{\partial}{\partial \eta} (\eta
Z), \eeq
the so-called virial equation of state (\ref{Zg}) and the
compressibility equation of state (\ref{chiS}) provide two
alternative routes for obtaining the compressibility factor
$Z(\eta)$ of a hard $d$-sphere fluid. Since all well-known
theoretical methods to obtain structural functions give approximate
results (with the exception of the exact solution for the
one-dimensional case $d=1$), one typically obtains two approximate
solutions, the compressibility factor from the virial route
\beq \label{Zv} Z_v(\eta) = 1 + 2^{d-1} \eta  g_c(\eta), \eeq
with $g_c(\eta)\equiv g(1^+)$, and the compressibility factor via
the compressibility route
\beq  \label{Zc} Z_{c}(\eta)=\int_0^1 {\dd x}{\chi^{-1}(\eta x)},
\eeq
with $\chi(\eta)$ given by Eq.\ (\ref{chiS}).

For hard-hypersphere fluids in arbitrary odd $d$ dimensions, the RFA
approach provides an analytical representation of the function
$G(s)$ defined by Eq.\ (\ref{GH}) and related to the structure
factor (\ref{SG}). In the simplest implementation of the RFA
approach, which is the so-called standard RFA and coincides with the
PY approximation, the function $G(s)$ can be written in the
{explicit} form\cite{RS07}
\beq
\label{Grfa}
G(s)= \frac{e^{- s}}{s^2} \frac{\sum_{j=0}^{n+1}
a_j s^j } { 1 + \an \eta \sum_{j=0}^{n+1} a_j\phi_{d-j}(s)} ,
\eeq
with
\beq
\label{aa} \an\equiv(-1)^{(d-1)/2} 2^{d-1} d!!,
\eeq
\beq
\label{phi} \phi_m(s) \equiv \frac 1{s^{m}} \left[
\sum_{j=0}^m \frac{(-s)^j}{j!} -e^{-s} \right].
\eeq
The coefficients $a_j$ are in general functions of the density.
Specifically, $a_0\equiv (d-2)!!$ and the quantities $a_j$ with
$j=1,\ldots,n+1$ are solutions of the following closed set of $n+1$
equations:
\beq  \label{ec1}
D_{2m+1}-\frac{a_{2m+1}}{(d-2)!!}+\sum_{j=0}^{m-1} \gamma_{2j}
D_{2(m-j)-1} = 0, \quad 0\leq m\leq n, \eeq
with the boundary condition
\beq \label{aj} \left. a_j \right|_{\eta=0} =
\frac{(2n+2-j)!}{2^{n+1-j}(n+1-j)! j!}. \eeq
The coefficients $D_l$ are linear combinations of the $\{a_j\}$
given by $D_0=1$ and
\beq \label{Dj} D_l= \frac{1}{l!} -\an\eta \sum_{m=0}^{n+1} a_m
\sum_{j=1}^l \frac{(-1)^{j+d-m}}{(j+d-m)!  (l-j)!}, \quad l\ge 1,
\eeq
and the quantities $\gamma_{2m}$ with $0\le m \le n$ are given in
terms of the coefficients $a_j$ by means of the recursion relation
\beq  \label{gammaj} \gamma_{2m}= \frac{{a}_{2m+2}}{(d-2)!!}
-D_{2m+2} -\sum_{j=0}^{m-1}\gamma_{2j} D_{2(m-j)},\quad 0\le m \le
n. \eeq
Here we have adopted the conventions $a_j=0$ if $j>n+1$ and
$\sum_{j=0}^m \cdots=0$ if $m<0$. In summary, when the
$\{\gamma_{2m}\}$ obtained from Eq.\ (\ref{gammaj}) are inserted
into Eq.\ \eqref{ec1}, ans use is made of Eq.\ (\ref{Dj}), one gets
 a closed set of $n+1=(d-1)/2$ algebraic
equations for $a_1$, $a_2$, \ldots, $a_{n+1}$. The number of
mathematical solutions (including complex ones) is $2^n=2^{(d-3)/2}$
and the physical solution is obtained as the one yielding the
correct low density limit given by Eq.\ (\ref{aj}).

The contact value of the rdf and the isothermal susceptibility
obtained by the PY theory (or, equivalently, by the standard RFA
method) are
\beq  \label{gc} g_{c}(\eta)= a_{n+1}
\left[1+\an\eta\sum_{j=0}^{n+1}\frac{(-1)^{d-j}}{(d-j)!} a_j,
\right]^{-1} \eeq
\beq
\label{chiRFA}
\chi(\eta)=1-2\lambda(d-2)!!\eta \Bigl( D_d
+\sum_{j=0}^{n}\gamma_{2j} D_{d-2-2j}\Bigr),
\eeq
with the factors $D_l$ and $\gamma_{2m}$ given by Eqs.\ (\ref{Dj})
and (\ref{gammaj}), respectively. With these results introduced into
Eqs. (\ref{Zv}) and (\ref{Zc}) one obtains the compressibility
factors $Z_v(\eta)$ and $Z_c(\eta)$ as derived within the PY theory
from the virial and compressibility routes, respectively.

\subsection{Virial expansions}

The virial expansion of the equation of state is an expansion in
powers of  the density or the packing fraction,
\beqa \label{virial} Z(\eta)=1+\sum_{j=2}^\infty B_j \rho^{j-1} =
1+\sum_{j=2}^\infty b_j \eta^{j-1}, \eeqa
and its range of validity is limited by the convergence properties
of the series. Notice that
\beq  b_j = B_j v_d^{1-j}. \eeq
Since $g_c(\eta=0)=1$, Eq. (\ref{Zg}) yields
\beq  \label{b2} b_2=2^{d-1}. \eeq
A more elaborated analysis of the virial equation shows that
$b_3=2^{2d-1} y_1(1)/y_1(0)$, where $y_1(r)$ represents the overlap
volume of two hyperspheres of unit radius and centers separated by a
distance $r$. In particular, for $d=\text{odd}$, one
finds\cite{RS07}
\beq  \label{b3} b_3=2^{2d-1} \left[1-\frac{(2n+3)!!}{2^{n+2}}
\sum_{j=0}^{n+1}\frac{(-4)^{-j}}{(2j+1)j!(n+1-j)!} \right]. \eeq
The application of the PY theory yields two virial expansions, one
for $Z_v(\eta)$ and another for $Z_c(\eta)$. The virial
coefficients in the virial route are given by
\beq \label{bjv}b_j^{(v)} = \frac{2^{d-1}}{(j-2)!} \left.
\partial_\eta^{j-2} g_c(\eta)\right|_{\eta=0}, \eeq
whereas in the compressibility route one has
\beq \label{bjc}b_j^{(c)} = \frac{1}{j!} \left.
\partial_\eta^{j-1}\chi^{-1}(\eta)\right|_{\eta=0}, \eeq
with $g_c(\eta)$ and $\chi(\eta)$ given by Eqs. (\ref{gc}) and
(\ref{chiRFA}), respectively, and where $\partial_\eta$ denotes the
derivative with respect to $\eta$. {In practice, what one does is to
solve Eq.\ \eqref{ec1} in a recursive way for the coefficients in
the density expansion of $a_1,\ldots,a_{n+1}$. The solutions are
exact rational numbers and from them one gets $b_j^{(v)}$ and
$b_j^{(c)}$ also as exact rational numbers.}

\section{Results}
\label{sect3}

In this section we present the results that follow from the
previous derivations. Three aspects will be analyzed. We first
deal with the virial coefficients. Then, we examine the issue of
the convergence properties of the virial series and finally we
compare the resulting compressibility factors with simulation
data.

\subsection{Virial coefficients}

\begin{table}
\caption{\label{ta.0} {Exact and PY values for the virial
coefficients $b_j/b_2^{j-1}$ for $4\leq j\leq 10$ and several
dimensionalities. The exact values are taken from Refs.\
\protect\onlinecite{CM03,CM06,BCW08}, while the PY values for hard
disks ($d=2$) are obtained from Ref.\
\protect\onlinecite{A-BKV08}}.}
\begin{ruledtabular}
\begin{tabular}{cccc}
 &Exact&PY (v)&PY (c)\\
 \hline
$d$ &  $b_4/b_2^3$ & $b_4^{(v)}/b_2^3$&$b_4^{(c)}/b_2^3$ \\
\hline
$2$&$   0.53223180$&$    0.5008 $&$        0.5378$\\
 $3$&$   0.2869495$&$     0.25$&$           0.296875$\\
 $5$&$   0.07597248$&$    0.04785156 $&$    0.08905029$\\
 $7$&$   0.00986495$&$   -0.007499695 $&$   0.021550816$\\
 $9 $&$ -0.008581$&$     -0.018590778 $&$   0.000373515$\\
 $11$&$  -0.011334 $&$    -0.016933400 $&$  -0.005004249$\\
 $13$&$-0.009523$&$ -0.012604546 $&$  -0.005256304$\\
 $15 $&$ -0.006934  $&$   -0.008614616 $&$-0.004146893$\\
 \hline
$d$ &  $b_5/b_2^4$ & $b_5^{(v)}/b_2^4$&$b_5^{(c)}/b_2^4$ \\
\hline
 $2$&$   0.33355604$&$    0.2948$&$        0.3433$\\
 $3$&$   0.110252$&$      0.0859375$&$     0.12109375$\\
 $5$&$   0.0129551  $&$   0.01294708 $&$   0.01638031$\\
 $7$&$   0.0070724  $&$   0.01235023 $&$   0.00511681$\\
 $9$&$   0.007439$&$     0.01172436 $&$   0.00404299$\\
 $11$&$   0.006176 $&$     0.00873948 $&$   0.00320718$\\
 $13$&$   0.004307 $&$0.00566100  $&$  0.00221818$\\
 $15$&$   0.002705 $&$     0.00337395$&$    0.00138891$\\
 \hline
 $d$ &  $b_6/b_2^5$ & $b_6^{(v)}/b_2^5$&$b_6^{(c)}/b_2^5$ \\
\hline
 $2 $&$  0.1988425 $&$    0.1667 $&$       0.2090$\\
 $3$&$   0.03888198$&$    0.02734375$&$    0.04492188$\\
 $5 $&$  0.0009815$&$    -0.00286007$&$    0.00145829$\\
 $7$&$  -0.0035121$&$    -0.00817700 $&$  -0.00186533$\\
 $9$&$  -0.004794 $&$     -0.00820391$&$   -0.00229461$\\
 $11$&$  -0.00395$&$      -0.00583856 $&$  -0.00179806$\\
 $13 $&$ -0.002580  $&$   -0.00347797 $&$  -0.00113940$\\
 $15$&$  -0.001472 $&$    -0.00186395$&$   -0.00063735$\\
\hline
 $d$ &  $b_7/b_2^6$ & $b_7^{(v)}/b_2^6$&$b_7^{(c)}/b_2^6$ \\
\hline
 $2$&$   0.1148728$&$     0.09194$&$       0.1233$\\
 $3$&$   0.01302354 $&$   0.00830078 $&$   0.015625$\\
 $5 $&$  0.0004162 $&$    0.00245853 $&$   0.00063127$\\
 $7 $&$  0.0025386  $&$   0.00655314 $&$   0.00138425$\\
 $9$&$0.003716$ &$0.00676771$&$0.00164806$\\
 \hline
\hline
 $d$ &  $b_8/b_2^7$ & $b_8^{(v)}/b_2^7$&$b_8^{(c)}/b_2^7$ \\
\hline
 $2 $&$  0.0649930 $&$    0.04988$&$       0.07107$\\
 $3 $&$  0.0041832$&$     0.00244140 $&$   0.00518799$\\
 $5$&$  -0.0001120 $&$   -0.00173717 $&$  -0.00022064$\\
 $7 $&$ -0.0019937 $&$   -0.00576280 $&$  -0.00107878$\\
 $9$&$-0.003222$&$-0.00621239$&$-0.00133333$\\
 \hline
 $d$ &  $b_9/b_2^8$ & $b_9^{(v)}/b_2^8$&$b_9^{(c)}/b_2^8$ \\
\hline
$ 2 $&$  0.0362193  $&$   0.02677$&$       0.04025$\\
 $3 $&$  0.0013094 $&$    0.00070190 $&$   0.00166321$\\
 $5 $&$  0.0000747 $&$    0.00134572 $&$   0.00017676$\\
 $7 $&$  0.0016869   $&$  0.00542528  $&$  0.00091218$\\
 $9$&$0.003029$&$0.00615464$&$0.00117921$\\
  \hline
 $d$ &  $b_{10}/b_2^9$ & $b_{10}^{(v)}/b_2^9$&$b_{10}^{(c)}/b_2^9$ \\
\hline
 $2 $&$  0.0199537  $&$   0.01424$&$       0.02248$\\
$ 3 $&$  0.0004035 $&$    0.00019836$&$    0.00051880$\\
 $5 $&$ -0.0000492   $&$ -0.00108546 $&$  -0.00012788$\\
 $7 $&$ -0.001514 $&$    -0.00537523$&$   -0.00081922$\\
  $9$&$-0.00306$&$-0.00645733$&$-0.00111652$\\
\end{tabular}
\end{ruledtabular}
%\footnotetext[1]{.}
\end{table}

\begin{figure}
\includegraphics[width=1.0\columnwidth]{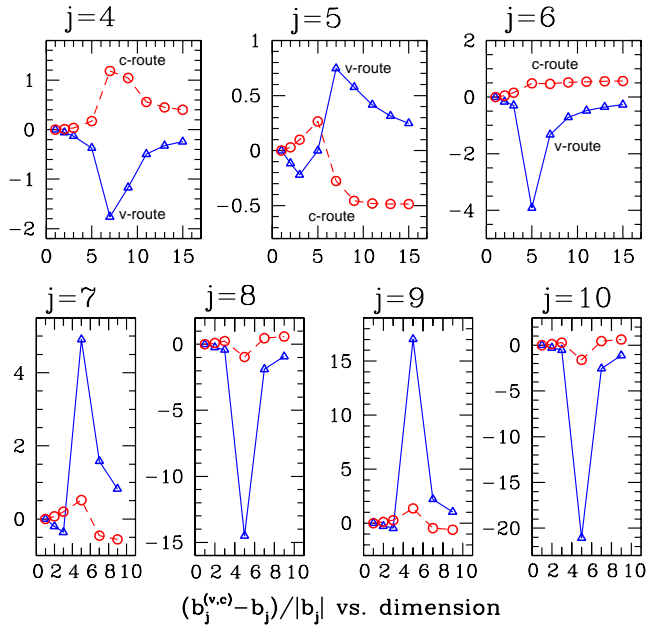}
\caption{{(Color online)} Normalized differences between the virial
coefficients $b_j^{(v)}$ (triangles) and $b_j^{(c)}$ (circles) and
the exact values\protect\cite{CM03,CM06,BCW08}  $b_j$, as functions
of the space dimension for $4\leq j \leq 10$. The data for $d=2$
have been {obtained from}  Ref.\ \protect\onlinecite{A-BKV08}. The
lines have been drawn to guide the eye.\label{f.bv_bc}}
\end{figure}

Because the PY theory is exact to first order in density, the virial
coefficients $b_2$ and $b_3$ [Eqs.\ (\ref{b2}) and (\ref{b3})] are
exactly reproduced by both routes. Higher virial coefficients are,
however, different for each route, {as shown in Table \ref{ta.0} for
$4\leq j\leq 10$, where the known exact values are also included for
the sake of comparison.} The normalized differences
$(b_j^{(v,c)}-b_j)/|b_j|$ between the approximate virial
coefficients $b_j^{(v)}$ and $b_j^{(c)}$ and the {exact
values\cite{CM03,CM06,BCW08}} $b_j$ are shown in Fig.\ \ref{f.bv_bc}
as functions of the space dimension for $4\leq j \leq 10$. As one
can see from the results in this figure, all known (exact)
coefficients in $d=2$ and $d=3$ lie between $b_j^{(c)}$ and
$b_j^{(v)}$ in the form $b_j^{(v)}<b_j<b_j^{(c)}$, but this is not
so in higher dimensions. A transition behavior seems to take place
at $d=5$ since in that case one has $b_j<b_j^{(c)}<b_j^{(v)}$ for
$j=7$ and 9, while $b_j>b_j^{(c)}>b_j^{(v)}$ for $j=8$ and 10. For
$d\geq 7$, however, the trends seem to be $b_j^{(v)}<b_j<b_j^{(c)}$
for $j=\text{even}\geq 4$ and $b_j^{(c)}<b_j<b_j^{(v)}$ for
$j=\text{odd}\geq 5$. The top panels of Fig.\ \ref{f.bv_bc} seem to
indicate that the relative deviations of the PY values with respect
to the exact ones tend to decrease and stabilize with increasing
$d$, especially in the case of $b_j^{(c)}$. Of course, a
confirmation of all these trends would require the knowledge of the
exact virial coefficients for higher orders $j$ and higher
dimensionalities $d$. The fact that $b_4^{(v)}<b_4<b_4^{(c)}$ for
all $d$ implies that $Z_v(\eta)<Z(\eta)<Z_c(\eta)$ for
asymptotically low densities. On the other hand, since both
$Z_v(\eta)$ and $Z_c(\eta)$ keep being finite {for any
$\eta<1$,\cite{RS07} i.e.,} even for densities higher than the
close-packing value $\eta_\text{cp}$,  it can be reasonably expected
that $Z(\eta)>Z_c(\eta)>Z_v(\eta)$ beyond a certain density,
although this possibly happens in the metastable region. A precursor
of this effect might be the relation $b_5^{(c)}<b_5<b_5^{(v)}$ for
$d\geq 7$.

\begin{figure}
{\includegraphics[width=1.0\columnwidth]{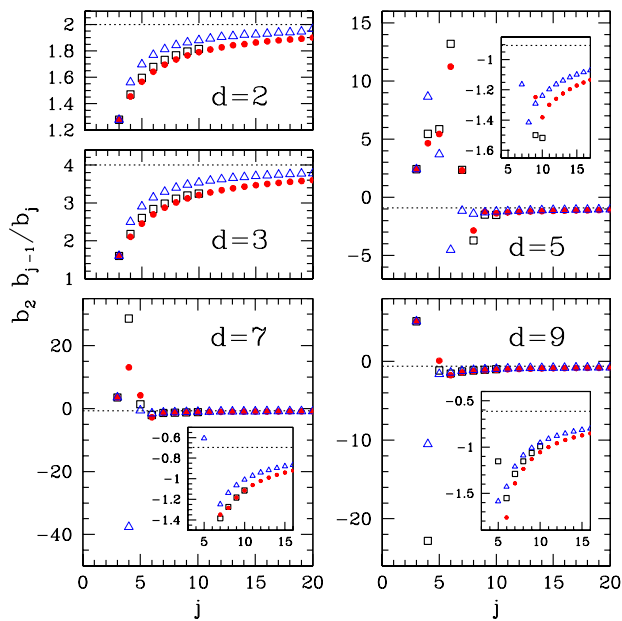}} \caption{{(Color
online)} Ratio $b_2 b_{j-1}/b_{j}$ as a function of $j$ for $3\leq j
\leq 20$ and $d=2$, 3, 5, 7, and $9$; $b_j^{(v)}$ (triangles),
$b_j^{(c)}$ (circles), and exact value\protect\cite{CM03,CM06,BCW08}
$b_j$ (squares). The dash-dotted line is the limiting value of this
ratio for both PY virial coefficients when $j \to \infty$. The
insets {in the cases $d=5$, 7 and $9$} show in more detail how the
limiting value is approached.
\label{f.bb_5}}
\end{figure}

The assessment of the performance of the PY virial coefficients with
respect to the exact results may also profit from a different
representation of the data. This is shown in the four panels of
Fig.\ \ref{f.bb_5}, where we have plotted the ratio $b_2
b_{j-1}/b_{j}$ as a function of $j$ for $3\leq j \leq 20$ (with both
$b_j^{(v)}$ and $b_j^{(c)}$  and with the values from Refs.\
\onlinecite{CM03,CM06,BCW08}) for $d=2$, 3, 5, 7, and $9$. As we
will discuss below, the magnitude of this ratio is related to the
radius of convergence of the virial series. In this instance, the
regularity of the results for the lower $j$'s observed for $d=2$ and
$d=3$ is lost in higher dimensions. It is interesting to note that
for $d=7$ and 9 the exact values of $b_2 b_{j-1}/b_j$ for the higher
$j$'s ($7\leq j\leq 10$) lie very close to the PY  values. {Whether
this good agreement  is accidental or not cannot be assessed before
exact values of $b_j$ for $j\geq 11$ and/or $d\geq 11$ are known.}

%
% =========================== figures

% =========================== end figures

\subsection{Radius of convergence of the virial series}

\begin{figure}
{\includegraphics[width=1.0\columnwidth]{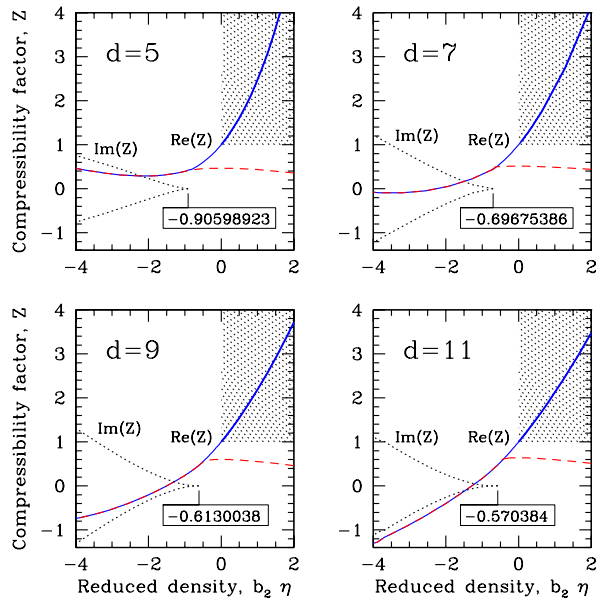}}
\caption{{(Color online)} Compressibility factor $Z_v(\eta)$ in the
physical domain of positive densities (thick solid line and shaded
region) and its analytic continuation for $\eta<0$ (thin solid line)
as a function of $b_2 \eta$ for $d=5$, 7, 9, and $11$. At the
negative value $\eta=-\eta_0$  $Z_v(\eta)$ merges with an unphysical
real root (dashed line) and both roots become a pair of complex
conjugates for $\eta<-\eta_0$  whose imaginary parts are represented
by the two dotted lines. {The boxed numbers are the corresponding
values of $-b_2\eta_0$.}\label{f.zz_b2rho}}
\end{figure}

\begin{figure}
{\includegraphics[width=1.0\columnwidth]{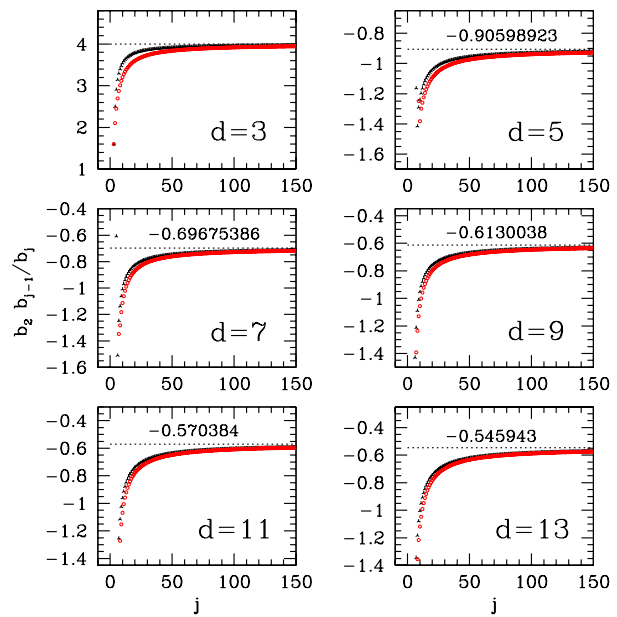}}
\caption{{(Color online)} Ratio $b_2 b_{j-1}/b_{j}$ as a function of
$j$ for $3\leq j \leq 150$ and  $d=3$, 5, 7, 9, 11, and $13$;
$b_j^{(v)}$ (triangles) and $b_j^{(c)}$ (circles). The dash-dotted
line is the limiting value of this ratio for both PY virial
coefficients when $j \to \infty$ and from such value one can
immediately get the radius of convergence of the virial
series.\label{f.bb_b2} }
\end{figure}

Now we turn to the question of the convergence properties of the
virial series. The radius of convergence of the virial series for
each dimension $d$, $\eta_0= \lim_{j\rightarrow
\infty}|b_{j-1}/b_j|$, is determined by the modulus of the
singularity of $Z(\eta)$ closest to the origin in the complex $\eta$
plane. {In order to inhibit the influence of $d$ on the
characteristic density values, we will sometimes choose $b_2\eta$
(rather than $\eta$ or $\rho\sigma^d$) to measure the density.} In
Fig.\ \ref{f.zz_b2rho} we display the {PY} compressibility factor
$Z_v(\eta)$ as a function of $b_2 \eta$ for $d=5$, 7, 9, and $11$.
In the figure, apart from showing $Z_v(\eta)$ in the physical domain
of positive densities (thick solid line, shaded region), we have
provided its analytic continuation to \textit{negative} values of
$\eta$. The thin solid line shows such a continuation. It turns out
that there exists a certain negative value $\eta=-\eta_0$ such that
$Z_v(\eta)$ keeps being real  in the interval $-\eta_0<\eta<0$.
However, at $\eta=-\eta_0$ $Z_v(\eta)$ merges with an unphysical
root (dashed line) and both roots become a pair of complex
conjugates for $\eta<-\eta_0$ (the two dotted lines represent their
imaginary parts). This shows that $Z_v(\eta)$, as well as
$Z_c(\eta)$, possesses a branch point at $\eta=-\eta_0$. This is the
singularity on the real axis closest to the origin. If no other
singularity lying in the complex plane is closer to the origin, then
$\eta_0$ is the radius of convergence of the series. Figure
\ref{f.bb_b2} provides the radius of convergence of the virial
series for $Z_v(\eta)$ and $Z_c(\eta)$, this time by representing
again $b_2 b_{j-1}^{(v,c)}/b_{j}^{(v,c)}$ as a function of $j$ for
$3\leq j \leq 150$ and $d=3$, 5, 7, 9, 11, and $13$. As {is
well-known}, the radius of convergence {predicted by the PY theory}
is $\eta_0=1$ for $d=3$. On the other hand, for $d\geq 5$ the radius
of convergence {is $\eta_0<1$ and} coincides with the value $\eta_0$
corresponding to the branch point on the negative real axis shown in
Fig.\ \ref{f.zz_b2rho}. The values of $\eta_0$, $b_2\eta_0$, and of
the \textit{scaled density per dimension}\cite{FP87}
$\widehat{\rho}_0=2\eta_0^{1/d}$  are shown in Table \ref{ta.1} for
odd dimensions in the interval $3\le d\le 13$. It can be observed
that the values of $\widehat{\rho}_0$ are consistent with the limit
$\widehat{\rho}_0\to 1$ as $d\to\infty$ conjectured by Frisch and
Percus.\cite{FP87} This agreement, along with the behavior observed
in {Fig.\ \ref{f.bb_5}}, supports the reliability of the radius of
convergence predicted by the PY theory, at least for high
dimensions.

%
% =========================== figures

% =========================== end figures

% =========================== table
\begin{table}
\caption{\label{ta.1} The radius of convergence $\eta_0$ of the
virial series in the PY theory, and the associated quantities
$b_2\eta_0$ and $\widehat{\rho}_0 = 2 \eta_0^{1/d}$ for fluids at
$d=3$, 5, 7, 9,11, and $13$.}
\begin{ruledtabular}
\begin{tabular}{cccc}
$d$ &  $\eta_0$ & $b_2\eta_0$&$\widehat{\rho}_0$ \\
\hline
 $3$ & $1$        &$4$      & $1.58740105$ \\
 $5$ & $0.056624327$ & $0.90598923$& $1.12623911$ \\
 $7$ & $0.010886779$ &$0.69675386$& $1.04854495$ \\
 $9$ & $0.002394546$ & $0.6130038$&$1.02289857$ \\
$11$ & $0.000557016$ & $0.570384$& $1.01204491$ \\
$13$ & $0.000133287$ & $0.545943$&$1.00678503$ \\
\end{tabular}
\end{ruledtabular}
%\footnotetext[1]{.}
\end{table}
% =========================== end table
%

\subsection{Compressibility factors}
\begin{figure}
{\includegraphics[width=1\columnwidth]{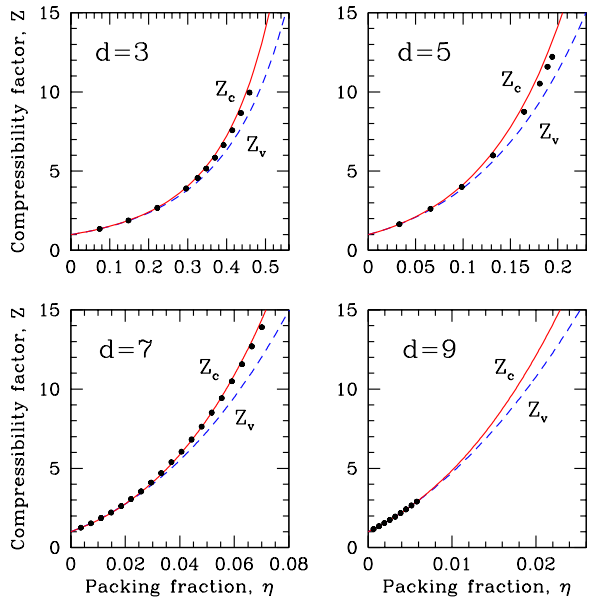}}
\caption{{(Color online)} Compressibility factors $Z_v(\eta)$
(dashed line) and $Z_c(\eta)$ (continuous line) of the PY theory as
functions of the packing fraction $\eta$ for $d=3$, 5, 7, and $9$,
and simulation data\protect\cite{LM90,RLdeHS04,BCW08,EW84,ER08}
(filled circles).
\label{f.zz_simul} }
\end{figure}

We now turn to the compressibility factor. In Fig.\ \ref{f.zz_simul}
we present a comparison between the PY values for $Z_v(\eta)$ and
$Z_c(\eta)$ and some of the available simulation
data\cite{EW84,MT84,LM90,RLdeHS04,L05,BW05,LB06,BW07,BCW08,ER08} for
various dimensions. It follows from this figure that, as $d$
increases, $Z_c(\eta)$ becomes a rather accurate approximation for
the simulation results. {Note, however, that the simulation data for
$d=9$ are restricted to the  density region where $Z_v(\eta)\simeq
Z_c(\eta)$.}

\begin{figure}
{\includegraphics[width=1\columnwidth]{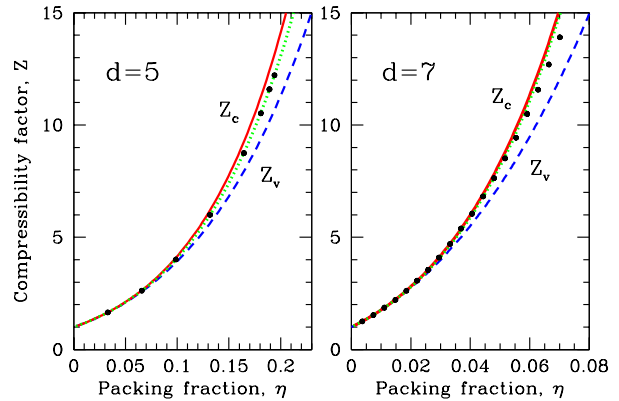}}
\caption{{(Color online) Truncated approximations $Z_v^*(\eta)$
(dashed line), $Z_c^*(\eta)$ (continuous line), and $Z^*(\eta)$
(dotted line) as functions of the packing fraction $\eta$ for $d=5$
and $7$, and simulation data}\protect\cite{LM90,RLdeHS04,ER08}
(filled circles).
\label{f.zz_simul2} }
\end{figure}

{Comparison between the density range in Fig.\ \ref{f.zz_simul} and
the values of $\eta_0$ tabulated in Table \ref{ta.1} shows that the
good agreement between $Z_c$ and $Z$ for $d=5$ and $d=7$ extends to
$\eta>\eta_0$, i.e., well beyond the radius of convergence of $Z_v$
and $Z_c$. This might cast doubts on the practical usefulness of the
virial coefficients to predict the equation of state of hard
hyperspheres in the fluid region with $\eta>\eta_0$. However, as the
following discussion shows,  we have
 observed that this is not the case. Let us denote by
$j^*$ the order of the virial coefficient preceding the first
negative coefficient, i.e., $b_j>0$ for $j\leq j^*$ but
$b_{j^*+1}<0$. Analogously one can define $j^*_v$ and $j^*_c$ in
connection with $b_j^{(v)}$ and $b_j^{(c)}$, respectively. For
instance, at $d=5$ one has (see Table \ref{ta.0}) $j^*=j_c^*=8$ and
$j^*_v=6$, while  $j^*=j_c^*=6$ and $j^*_v=4$ at $d=7$. Let us now
define the \emph{truncated} virial series}
\beq
Z^*(\eta)=1+\sum_{j=2}^{j^*}b_j \eta^{j-1}, \label{Z^*}
\eeq
{with similar definitions of $Z_v^*(\eta)$ and $Z_c^*(\eta)$. Since
the first neglected term is negative, it follows that, at least for
low densities, $Z^*(\eta)$, $Z^*_v(\eta)$, and $Z^*_c(\eta)$ are
upper bounds of $Z(\eta)$, $Z_v(\eta)$, and $Z_c(\eta)$,
respectively. Furthermore, we have checked  (for $d=5$, 7, 9, 11,
and 13) that $Z^*_v(\eta)$ and $Z^*_c(\eta)$ are excellent estimates
of $Z_v(\eta)$ and $Z_c(\eta)$, respectively, much better in the
region $\eta>\eta_0$ than any other truncated virial series. Does
this property extend to $Z^*(\eta)$ as well? Comparison between
$Z^*(\eta)$ and simulation data of $Z(\eta)$ for $d=5$ and 7
indicate that this is indeed the case.  In Fig.\ \ref{f.zz_simul2}
we have represented the approximate compressibility factors
$Z^*_v(\eta)$, $Z^*_c(\eta)$, and $Z^*(\eta)$  for $d=5$ and $7$,
and compared such approximations with simulation
values.\cite{LM90,RLdeHS04,ER08} Again the superiority of the
approximate PY compressibility factor obtained via the
compressibility route over the one obtained via the virial route is
apparent. More interesting, however, is the fact that if one
considers the similar approximation $Z^*(\eta)$ (i.e., using the
exact virial coefficients), the agreement between this approximation
and the simulation data is strikingly good over the whole fluid
phase region where these data are available, especially in the case
$d=5$. In fact, since as already pointed out, the PY theory leads to
the exact (positive) $b_2$ and $b_3$ (irrespective of whether one
takes the virial or the compressibility routes), and since for $d
\geq 9$ both the exact fourth virial coefficient and the PY
$b_4^{(v)}$ are negative, it turns out that  for these dimensions
$j^*=j_v^*=3$ and  the truncated expansions
$Z^*(\eta)=Z_v^*(\eta)=1+b_2\eta+b_3\eta^2$  coincide. Furtermore,
since the PY coefficient $b_4^{(c)}$ also becomes negative for $d
\geq 11$, all three truncated expansions for the compressibility
factor become the same for $d \geq 11$.}

%
% =========================== figures

% ===========================  end figures
%
\section{Discussion}
\label{sect4}

The preceding results lend themselves to further consideration. One
might have reasonably wondered whether the trend observed for all
the known virial coefficients $b_j$ both in $d=2$ and $d=3$, namely
the fact that they are bracketed by $b_j^{(v)}$ and $b_j^{(c)}$,
would remain valid for all virial coefficients in these dimensions
and also hold for the higher dimensions. The following reasoning
indicates that for any $d > 1$ there must be at least one virial
coefficient that does not comply with the above trend. Since both
$Z_v(\eta)$ and $Z_c(\eta)$ only diverge for $\eta=1$ and the true
$Z(\eta)$ must have a divergence at the close-packing fraction
$\eta_\text{cp}<1$, then at a smaller packing fraction (although
possibly in the metastable fluid region), $Z(\eta)$ must lie above
$Z_v(\eta)$ and $Z_c(\eta)$. This crossing should manifest itself in
the behavior of the virial coefficients even if the radius of
convergence of the virial series is smaller (or very much smaller)
than the packing fraction at which the crossing occurs. According to
our findings, the bracketing {$b_j^{(v)}<b_j<b_j^{(c)}$} is lost
already for $b_5$ at $d=7${, where one has
$b_j^{(c)}<b_j<b_j^{(v)}$,} and it seems that this will happen for
all higher odd $j$'s but not for even ones. We also find that the
relative deviations of the PY virial coefficients with respect to
the true coefficients tend to stabilize with increasing
dimensionality.

Concerning the radius of convergence of the virial series, two key
aspects should be pointed out. The first one is that, as our present
results clearly confirm, in the PY theory and for $d \geq 5$ the
singularity closest to the origin is a branch point located on the
negative real axis and corresponds to a packing fraction $\eta_0$
much smaller than the one delimiting the fluid phase region. The
second observation is that the PY values for the ratio $b_2
b_{j-1}/b_j$ for the higher $j$'s lie very close to the exact
values. Further, the resulting scaled density per dimension
$\widehat{\rho}_0$ is wholly consistent with the limiting value of
$1$ for $d \to \infty$. Hence, on the one hand the scenario of an
alternating series for the true virial series by Clisby and
McCoy\cite{CM06} {(perhaps even for $d=3$)} is reinforced and, on
the other, one may conjecture that its radius of convergence will
also be close to the one in the PY theory. {In connection with the
latter point it is instructive to compare the PY values of the
radius of convergence $\eta_0$ with known bounds. In 1964, Lebowitz
and Penrose\cite{LP64} derived the lower bound $b_2\eta_0\geq
0.07238$, which is clearly verified by the values of Table
\ref{ta.1}. They also showed that no phase transitions exist for
values of $b_2\eta$ less than $1/2(1+e)\simeq 0.135$, which is again
consistent with the PY predictions for $\eta_0$ and the
scenario\cite{CM06} of a leading singularity on the positive real
axis at the freezing density $\eta_f>\eta_0$. Recently, Clisby and
McCoy\cite{CM03} estimated the value $b_2\eta=0.5$ for the radius of
convergence of the sum of the Ree--Hoover diagrams. It is quite
interesting to notice that the values of $b_2\eta_0$ predicted by
the PY approximation (see Table \ref{ta.1}) seem to tend from above
precisely to 0.5, i.e., $\lim_{d\to\infty}b_2\eta_0=0.5$.}

As a final point, it is worth mentioning that the regularity
observed for the ratio $(Z_v-Z)/(Z_c-Z_v)$ in $d=3$ and $d=5$, which
leads to the Carnahan--Starling equation of state\cite{CS69} in the
former case and something similar in the latter,\cite{S00} seems not
to hold for $d=7$, even if the low density values (which are
subjected to great uncertainties) are ignored. The recent simulation
data of Bishop \textit{et al.}\cite{BCW08} are also for low density
and with error bars comparable to $Z_c-Z_v$ so that further
conclusions on this matter are precluded at this stage.

\begin{acknowledgments}

R. D. Rohrmann acknowledges the financial support of the SeCyT
(Argentina)  through Grant 114/07. M. Robles and M. L\'opez de Haro
acknowledge the partial financial support of DGAPA-UNAM under
project IN-109408. A. Santos acknowledges the financial support of
the Ministerio de Educaci\'on y Ciencia (Spain) through Grant No.\
FIS2007--60977 (partially financed by FEDER funds) and by the Junta
de Extremadura through Grant No.\ GRU08069.

\end{acknowledgments}

\end{document}